Anthropic calculation of the velocity and acceleration of the space-time

Miroslaw Kozlowski[1,3] and Janina Marciak-Kozlowska[2]

[1,3]Institute of Experimental Physics, Warsaw University Warsaw Poland,

[2]Institute of Electron Technology Warsaw Poland

Abstract

In this paper considering quantum heat transport equation (QHT) formulated in our earlier papers the temperature for universes with $G < 0$ is calculated. As the solution of complex QHT (Schrodinger type equation), the temperature is complex also. We argue that due to anthropic limitation of the observer ImT(r,t)=0 From this condition the discretization of space-time radius $R$, velocity of the universe expansion $v$, Hubble parameter $H$ and acceleration of the expansion $a$ are calculated. The agreement with observational data for our Universe is quite good.

1 Introduction

In the recent years the growing interest for the source of Universe expansion is observed. After the work of Supernova detecting groups the consensus for the acceleration of the moving of the space time is established [1,2].

In this paper we follow of idea of the repulsive gravity as the source of the space-time expansion. We will study the influence of the repulsive gravity ($G < 0$) on the temperature field in the universe. To that aim we will apply the quantum hyperbolic heat transfer equation (QHT) formulated in our earlier papers [3,4].

When substitution $G \rightarrow -G$ is performed in QHT the Schrödinger type equation is obtained for the temperature field. In this paper the solution of QHT will be obtained. The resulting temperature is a complex function of space and time. We argue that because of the anthropic limitation of the observers it is quite reasonable to assume Im$T$=0. From this anthropic condition the discretization of the space radius $R=[(4N\pi+3\pi)L_P]^{1/2}(ct)^{1/2}$, velocity of expansion $v=(\pi/4)^{1/2}((N+3/4)/M)^{1/2}c$ and acceleration of expansion $a=-[1/2](\pi/4)^{1/2}((N+3/4)^{1/2}/M^{3/2})(c^7/((h/2\pi)G))^{1/2}$ are obtained.

2 Model

In papers [3,4] the quantum heat transport equation in a Planck Era was formulated:

$$\tau \frac{\partial^2 T}{\partial t^2} + \frac{\partial T}{\partial t} = \frac{(\hbar/2\pi)}{M_P} \nabla^2 T. \quad (1)$$

In equation (1) $\tau = ((\hbar/2\pi) G)/c^5)^{1/2}$ is the relaxation time, $M_P = (((\hbar/2\pi) c)/G)^{1/2}$ is the mass of the Planck particle, $(\hbar/2\pi)$, $c$ are the Planck constant and light velocity respectively and $G$ is the gravitational constant. The crucial role played by gravity (represented by $G$ in formula (1)) in a Planck Era was investigated in paper [4].

For a long time the question whether, or not the fundamental constant of nature $G$ vary with time has been a question of considerable interest. Since P. A. M. Dirac [5] suggested that the gravitational force may be weakening with the expansion of the Universe, a variable $G$ is expected in theories such as the Brans-Dicke scalar-tensor theory and its extension [6,7]. Recently the problem of the varying $G$ received renewed attention in the context of extended inflation cosmology [8].

It is now known, that the spin of a field (electromagnetic, gravity) is related to the nature of the force: fields with odd-integer spins can produce both attractive and repulsive forces; those with even-integer spins such as scalar and tensor fields produce a purely attractive force. Maxwell's electrodynamics, for instance can be described as a spin one field. The force from this field is attractive between oppositely charged particles and repulsive between similarly charged particles.

The integer spin particles in gravity theory are like the graviton, mediators of forces and would generate the new effects. Both the graviscalar and the graviphoton are expected to have the rest mass and so their range will be finite rather than infinite. Moreover, the graviscalar will produce only attraction, whereas the graviphoton effect will depend on whether the interacting particles are alike or different. Between matter and matter (or antimatter and antimatter) the graviphoton will produce repulsion. The existence of repulsive gravity forces can to some extent explains the early expansion of the Universe [5].

In this paper we will describe the influence of the repulsion gravity on the quantum thermal processes in the universe. To that aim we put in equation (1) $G \rightarrow -G$. In that case the new equation is obtained, viz.

$$i(\hbar/2\pi) \frac{\partial T}{\partial t} = \left( \frac{(\hbar/2\pi)^3 |G|}{c^5} \right)^{1/2} \frac{\partial^2 T}{\partial t^2} - \left( \frac{(\hbar/2\pi)^3 |G|}{c} \right)^{1/2} \nabla^2 T. \quad (2)$$

For the investigation of the structure of equation (2) we put:

$$\frac{(\hbar/2\pi)^2}{2m} = \left( \frac{(\hbar/2\pi)^3 |G|}{c} \right)^{1/2} \quad (3)$$

and obtains

$$m = \frac{1}{2} M_P$$

with new form of the equation (2)

$$i(\hbar/2\pi) \frac{\partial T}{\partial t} = \left( \frac{(\hbar/2\pi)^3 |G|}{c^5} \right)^{1/2} \frac{\partial^2 T}{\partial t^2} - \frac{(\hbar/2\pi)^2}{2m} \nabla^2 T. \tag{4}$$

Equation (4) is the quantum telegraph equation discussed in paper [4]. To clarify the physical nature of the solution of equation (4) we will discuss the diffusion approximation, *i.e.* we omit the second time derivative in equation (4) and obtain

$$i(\hbar/2\pi) \frac{\partial T}{\partial t} = -\frac{(\hbar/2\pi)^2}{2m} \nabla^2 T. \tag{5}$$

Equation (5) is the Schrödinger type equation for the temperature field in a universes with $G < 0$.

Both equation (5) and diffusion equation:

$$\frac{\partial T}{\partial t} = \frac{(\hbar/2\pi)^2}{2m} \nabla^2 T \tag{6}$$

are parabolic and require the same boundary and initial conditions in order to be ``well posed''.

The diffusion equation (6) has the propagator [10]:

$$T_D(\vec{R}, \Theta) = \frac{1}{(4\pi D \Theta)^{3/2}} \exp\left[ -\frac{R^2}{2\pi(\hbar/2\pi) \Theta} \right], \tag{7}$$

where

$$\vec{R} = \vec{r} - \vec{r'}, \qquad \Theta = t - t'.$$

For equation (5) the propagator is:

$$T_s(\vec{R}, \Theta) = \left(\frac{M_P}{2\pi(\hbar/2\pi)\Theta}\right)^{3/2} \exp\left[-\frac{3\pi i}{4}\right] \cdot \exp\left[\frac{iM_P R^2}{2\pi(\hbar/2\pi)\Theta}\right] \qquad (8)$$

with initial condition $T_s(\vec{R}, 0) = \delta(\vec{R})$.

In equation (8) $T_s(\vec{R}, \Theta)$ is the complex function of $\vec{R}$ and $\Theta$. For anthropic observers only the real part of $T$ is detectable, so in our description of universe we put:

$$\mathrm{Im} T(\vec{R}, \Theta) = 0. \qquad (9)$$

The condition (9) can be written as (bearing in mind formula (8)):

$$\sin\left[-\frac{3\pi}{4} + \left(\frac{R}{L_P}\right)^2 \frac{1}{4\tilde{\Theta}}\right] = 0, \qquad (10)$$

where $L_P = \tau_P c$ and $\tilde{\Theta} = \Theta/\tau_P$. Formula (10) describes the discretization of $R$

$$R_N = [(4N\pi + 3\pi)L_P]^{1/2}(tc)^{1/2}, \qquad (11)$$
$$N = 0, 1, 2, 3\ldots$$

In fact from formula (11) the Hubble law can be derived

$$\dot{R}_N = H = \frac{1}{2t}, \quad \text{independent of } N. \qquad (12)$$

In the subsequent we will consider $R$ (11), as the space-time radius of the $N-$ universe with ``atomic unit'' of space $L_P$.

It is well known that idea of discrete structure of time can be applied to the ``flow'' of time. The idea that time has ``atomic'' structure or is not infinitely divisible, has only recently come to the fore as a daring and sophisticated hypothetical concomitant of recent investigations in the physics elementary particles and astrophysics. Yet in the Middle Ages the atomicity of time was maintained by various thinkers, notably by Maimonides [11]. In the most celebrated of his works: *The Guide for perplexed* he wrote: *Time is composed of time-atoms, i.e. of many parts, which on account of their short duration cannot be divided.* The theory of Maimonides was also held by Descartes [12].

The shortest unit of time, atom of time is named *chronon* [13]. Modern speculations concerning the *chronon* have often be related to the idea of the smallest natural length is $L_P$. If this is divided by velocity of light it gives the Planck time $\tau_P=10^{-43}$ s, *i.e. the chronon* is equal $\tau_P$. In that case the time $t$ can be defined as

$$t = M\tau_P, \qquad M=0, 1, 2, ... \tag{13}$$

Considering formulae (8) and (13) the space-time radius can be written as

$$R(M, N) = (\pi)^{1/2} M^{1/2} \left( N + \frac{3}{4} \right)^{1/2} L_P, \qquad M, N = 0, 1, 2, 3, \ldots \tag{14}$$

Formula (14) describes the discrete structure of space-time. As the $R(M, N)$ is time dependent, we can calculate the velocity, $v = dR/dt$, *i.e.* the velocity of the expansion of space-time

$$v = \left( \frac{\pi}{4} \right)^{1/2} \left( \frac{N+3/4}{M} \right)^{1/2} c, \tag{15}$$

where $c$ is the light velocity. We define the acceleration of the expansion of the space-time

$$a = \frac{dv}{dt} = -\frac{1}{2} \left( \frac{\pi}{4} \right)^{1/2} \frac{(N+3/4)^{1/2}}{M^{3/2}} \frac{c}{\tau_P}. \tag{16}$$

Considering formula (16) it is quite natural to define Planck acceleration:

$$A_P = \frac{c}{\tau_P} = \left( \frac{c^7}{(h/2\pi)G} \right)^{1/2} = 10^{51} \text{ ms}^{-2} \tag{17}$$

and formula (16) can be written as

$$a = -\frac{1}{2} \left( \frac{\pi}{4} \right)^{1/2} \frac{(N+3/4)^{1/2}}{M^{3/2}} \left( \frac{c^7}{(h/2\pi)G} \right)^{1/2}. \tag{18}$$

In table I the numerical values for *R*, *v* and *a* are presented. It is quite interesting that for *N*, *M*→∞ the expansion velocity
*v* < *c* in complete accord with relativistic description. Moreover for *N*, *M* >> 1 the *v* is relatively constant *v* ~ 0.88 *c*. >From formulae (11) and (15) the Hubble parameter *H*, and the age of our Universe can be calculated

$$v = HR, \quad H = \frac{1}{2M\tau_P} = 5 \cdot 10^{-18} \text{ s}^{-1},$$
$$T = 2M\tau_P = 2 \cdot 10^{17} \text{ s} \sim 10^{10} \text{ years}, \quad (19)$$

which is in quite good agreement with recent measurement [15,16,17].

In figs. 1(a),(b) the velocity and acceleration as the function of $L(L_P)$ and $T(T_P)$ are presented and in figs. 2(a),(b) the presented day radiuses for *N*, $M=10^{60}$ are presented.

## 3 Concluding remarks

In this paper following the QHT the discrete structure of the space time is investigated. Assuming anthropic condition Im$T([(r)\vec], t)=0$ the discretization of space-time is evaluated. The formulae for discrete radius $R(N, M)$, velocity $v(N, M)$ and acceleration are obtained. It is shown that numerical values $R(N, M)=10^{60}$ $L_P$, $v(N, M)=0.88c$ and $a(N, M)=4.43 \cdot 10^{-60}$ $A_P$ for $N=M=10^{60}$ are in good agreement with the observational data of our Universe.

Table I: *Radius, velocity and acceleration for N, M-universes*

| N, M | R[m] | v[m/s] | a[m/s$^2$] |
|---|---|---|---|
| $10^{20}$ | $1.77 \cdot 10^{-15}$ | $2.66 \cdot 10^{8}$ | $-1.32 \cdot 10^{31}$ |
| $10^{60}$ | $1.77 \cdot 10^{25}$ (*) | $2.66 \cdot 10^{8}$ (*) | $-1.32 \cdot 10^{-10}$ (**) |
| $10^{80}$ | $1.77 \cdot 10^{45}$ | $2.66 \cdot 10^{8}$ | $-1.32 \cdot 10^{-29}$ |

(*)Spergel D. N. *at al.* [16];

(**)Anderson J. D. *at al.* [17] Radio metric data from Pionier 10/11, Galileo and Ulysses Data indicate and apparent anomalous, constant, acceleration acting on the spacecraft with a magnitude $\sim 8.5 \cdot 10^{-10}$ m/s$^2$.

References

[1] Riess A. G. *et al.*, *Astron. J.*, **116** (1998) 1009.

[2] Starkman G. *et al.*, *Phys. Rev. Lett.*, **83** (1999) 1510.

[3] Marciak-Kozlowska J., *Found. Phys. Lett.*, **10** (1997) 295.

[4] Kozlowski M. and Marciak-Koz³owska J., *Found. Phys. Lett.*, **10** (1997) 599.

[5] Dirac P. A. M., *Nature* (London), **139** (1937) 323.

[6] Damour T., Gibbons G. W. and Gundach C., *Phys. Rev. Lett.*, **64** (1990) 123.

[7] Damour T. and Esposito Farèse G. Esposito, *Classical Quantum Gravity*, **9** (1992) 2093.

[8] Steinhard P. J., *Phys. Rev. Lett.*, **62** (1989) 376.

[9] Gasperini M., *Gen. Rel. Grav.*, **30** (1998) 1703.

[10] Barton G., in *Elements of Green's functions and propagation*, Oxford Science Publications, Clarendon Press, Oxford, 1995, p. 222.

[11] Maimonides M. (1190), in *The guide for the perplexed* (transl. M. Friedlander, 1904) p. 121 George Routledge, London.

[12] Descartes R., *Meditions on the first philosophy*. In A discourse on method etc. (transl. A. D. Lindsay 1912) pp. 107-108 Deut, London.

[13] Whitrow G. J., in *The natural philosophy of time*, Second Edition, edited by Oxford Science Publications, 1990, p. 204.

[14] Guth A. H., in *The Inflationary Universe: The quest for a New Theory of Cosmic Origins*, edited by Addison-Wesley, New York 1977; Guth A. H., PNAS **90** (1993) 4871.

[15] Cayrel R. *et al.*, *Nature*, **409** (2001) 691.

[16] Spergel D. N. *et al.*, PNAS **94** (1997) 6579.

[17] Anderson J. D. *et al.*, *Phys. Rev. Lett.*, **81** (1998) 2858.